\documentclass{article}
\usepackage{spconf,amsmath,graphicx}

\usepackage{enumitem}
\setlist{nosep, leftmargin=14pt}

\usepackage{mwe} % to get dummy images

\usepackage{amsmath,amsfonts,amssymb}
\usepackage{graphicx}
\usepackage[colorlinks=true, allcolors=blue]{hyperref}
\usepackage{mathtools}
\usepackage{bm}

\DeclarePairedDelimiter\floor{\lfloor}{\rfloor}

\usepackage{microtype}

\usepackage{array, makecell}
\usepackage{boldline}
\usepackage{colortbl}
\usepackage{color}
\definecolor{Gray}{gray}{0.9}
\usepackage{xcolor}
\title{From Registration Uncertainty to Segmentation Uncertainty}
\address{Johns Hopkins University, Baltimore, MD, USA}
\begin{document}
%\ninept
%
\maketitle
\begin{abstract}
Understanding the uncertainty inherent in deep learning-based image registration models has been an ongoing area of research. Existing methods have been developed to quantify both transformation and appearance uncertainties related to the registration process, elucidating areas where the model may exhibit ambiguity regarding the generated deformation.
However, our study reveals that neither uncertainty effectively estimates the potential errors when the registration model is used for label propagation. Here, we propose a novel framework to concurrently estimate both the epistemic and aleatoric segmentation uncertainties for image registration. To this end, we implement a compact deep neural network~(DNN) designed to transform the appearance discrepancy in the warping into aleatoric segmentation uncertainty by minimizing a negative log-likelihood loss function. Furthermore, we present epistemic segmentation uncertainty within the label propagation process as the entropy of the propagated labels. By introducing segmentation uncertainty along with existing methods for estimating registration uncertainty, we offer vital insights into the potential uncertainties at different stages of image registration. We validated our proposed framework using publicly available datasets, and the results prove that the segmentation uncertainties estimated with the proposed method correlate well with errors in label propagation, all while achieving superior registration performance. Code is available at \url{https://bit.ly/42VOZER}.
\end{abstract}
\begin{keywords}
Image registration, Registration uncertainty, Segmentation uncertainty
\end{keywords}
\section{INTRODUCTION}
\label{sec:intro}
%Deformable image registration (DIR) is pivotal in medical image analysis and serves as a fundamental component for many medical imaging applications. Traditional DIR methodologies necessitate optimization for each image pair, a process that can be computationally intensive and time-consuming. However, recent advancements in deep learning have shown promise in not only expediting the image registration process but also in enhancing the accuracy of registration when compared to traditional approaches~\cite{balakrishnan2019voxelmorph, chen2022transmorph}. These methods optimize a global objective function during a training phase, and then, during the testing phase, the registration is achieved through a forward pass. This bypasses the need for optimization on a per-image basis, making these methods fast. 

%Evaluating the uncertainty of a medical image analysis model is crucial as it offers physicians a measure of the model's prediction confidence. This, in turn, can assist them in making more well-informed clinical decisions or devising treatment plans. However, most deep neural networks (DNNs) used in medical imaging produce deterministic results that are presumed to be precise, but this is frequently not the case. Hence, methods such as 

Bayesian deep learning has been successfully employed for various medical imaging applications, including segmentation~\cite{eaton2018towards, jungo2019assessing} and registration~\cite{yang2017quicksilver, chen2022transmorph}, to facilitate the estimation of predictive uncertainty.
Generally, learning-based image registration algorithms consider two types of uncertainties---transformation and appearance uncertainties~\cite{chen2023survey}.
The former is reflective of the uncertainty in the deformation space, which tends to be larger when the registration model struggles to establish specific correspondences, such as registering regions with piece-wise constant intensity.
The latter, however, is typically premised on the belief that a high image similarity leads to accurate registration.
As such, this uncertainty would be considerable when appearance disparities exist between the warped and fixed images.
In applying registration models for image segmentation tasks (e.g., atlas-based image segmentation), the anatomical label of the moving image is propagated to the fixed image via the predicted deformation field.
Understanding the interconnection between the registration uncertainty and segmentation uncertainty is crucial, as leveraging the former could enhance the segmentation accuracy~\cite{simpson2011probabilistic}.
However, registration uncertainty cannot be directly interpreted as segmentation uncertainty.
% In such tasks, it becomes equally crucial to comprehend the uncertainty associated with the propagated label map. Although the uncertainties in registration and segmentation are interlinked, and the former may potentially enhance segmentation accuracy~\cite{simpson2011probabilistic}, registration uncertainty cannot be directly interpreted as segmentation uncertainty.
Specifically, transformation uncertainty usually underestimates the intensity misalignment between images, but this misalignment may be linked to errors in label propagation. On the other hand, appearance uncertainty may exhibit excessive sensitivity to image noise, but the noise contributions should not be regarded as part of the uncertainty in the label propagation, as long as the anatomical regions of the warped and fixed image register accurately.

In this paper, we propose to bridge the gap between registration uncertainty and segmentation uncertainty in learning-based image registration. Specifically, we propose a deep neural network~(DNN) that is conditioned on the appearance differences between the warped and fixed image to estimate the uncertainty in propagating the anatomical labels. The proposed method can estimate the aleatoric segmentation uncertainty without necessitating the actual anatomical label map at test time. Additionally, if anatomical labels are provided, the epistemic segmentation uncertainty can also be determined. The proposed method was evaluated on publicly available datasets, with favorable results corroborating its effectiveness.

\begin{figure*}
\centering
\includegraphics[width=0.6\textwidth]{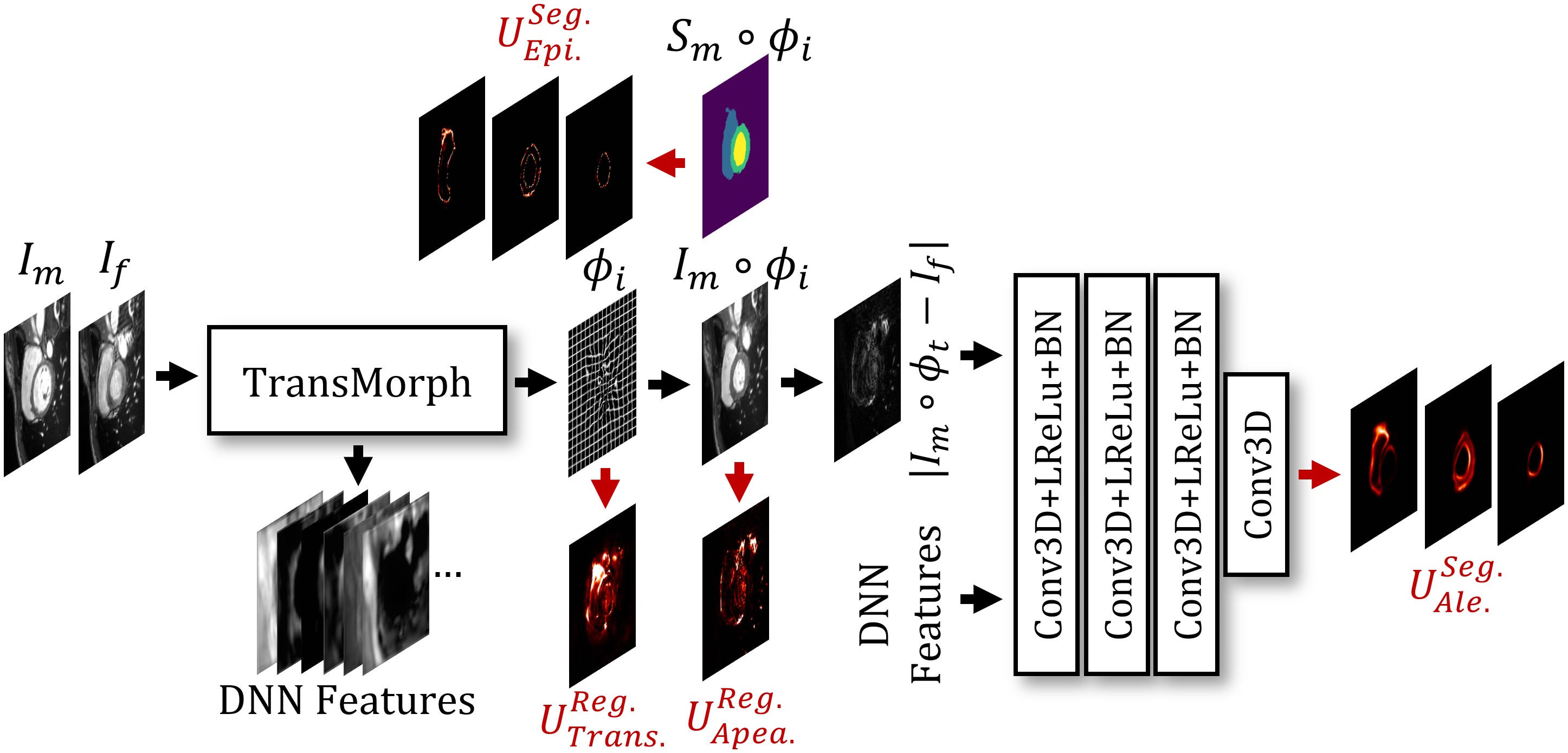}
\caption{The overall framework of estimating registration and segmentation uncertainty for DNN-based image registration.} \label{fig:framework}
\end{figure*}

\section{METHODS}
Figure~\ref{fig:framework} shows the overall framework of the proposed method. 
Let $I_f,I_m:\Omega\to\mathbb{R}$ be the fixed and moving images, respectively, which are defined over a subset of a 3-dimensional spatial domain, $\Omega \in \mathbb{R}^3$.
The registration network generates a deformation field $\phi$ that warps $I_m$ to $I_f$.
This network is built on the basis of our previously proposed Transformer-based registration network, TransMorph~\cite{chen2022transmorph, chen2022unsupervised}, which includes a Bayesian variant that integrates Bayesian deep learning by employing Monte Carlo dropout~\cite{gal2016dropout}.
From Chen~\textit{et al.}~\cite{chen2022transmorph}, we estimate \textit{transformation uncertainty} as the variance of the deformation field $\phi=\{\phi_i\ |\ i\in \{1, 2, \dots, T\}\}$, where $T$ is the sampling times. We also estimate the well-calibrated \textit{appearance uncertainty} as the mean squared difference between the warped image $I_m\circ\phi$ and the fixed image $I_f$.
Formally, these can be represented as:

\begin{equation}
\begin{split}
    U^{Reg.}_{Trans.} &= \frac{1}{T}\sum_{i=1}^T\left(\phi_i-\frac{1}{T}\sum_{i=1}^T\phi_i\right)^2,\\
    U^{Reg.}_{Appea.} &= \frac{1}{T}\sum_{i=1}^T(I_m\circ\phi_i-I_f)^2.
\end{split}
\end{equation}

It should be noted that $U^{Reg.}_{Trans.}$ defined here considers only the \textit{epistemic transformation uncertainty}, denoting the DNN's uncertainty in predicting the deformation. 
The \textit{aleatoric transformation uncertainty} might be estimated through probabilistic modeling of the deformation field~\cite{dalca2019unsupervised}, albeit this consideration is beyond the scope of this paper.
Conversely, $U^{Reg.}_{Apea.}$, defined above, is well-calibrated in the context of predictive error determined by the mean squared error~(MSE), encapsulating both the epistemic and aleatoric appearance uncertainty~\cite{chen2022transmorph}.
In the subsequent subsections, we detail the methodology for estimating both the \textit{epistemic} and \textit{aleatoric segmentation uncertainty} associated with label propagation during the image registration process.
\begin{figure*}[t]
\centering
\includegraphics[width=0.6\textwidth]{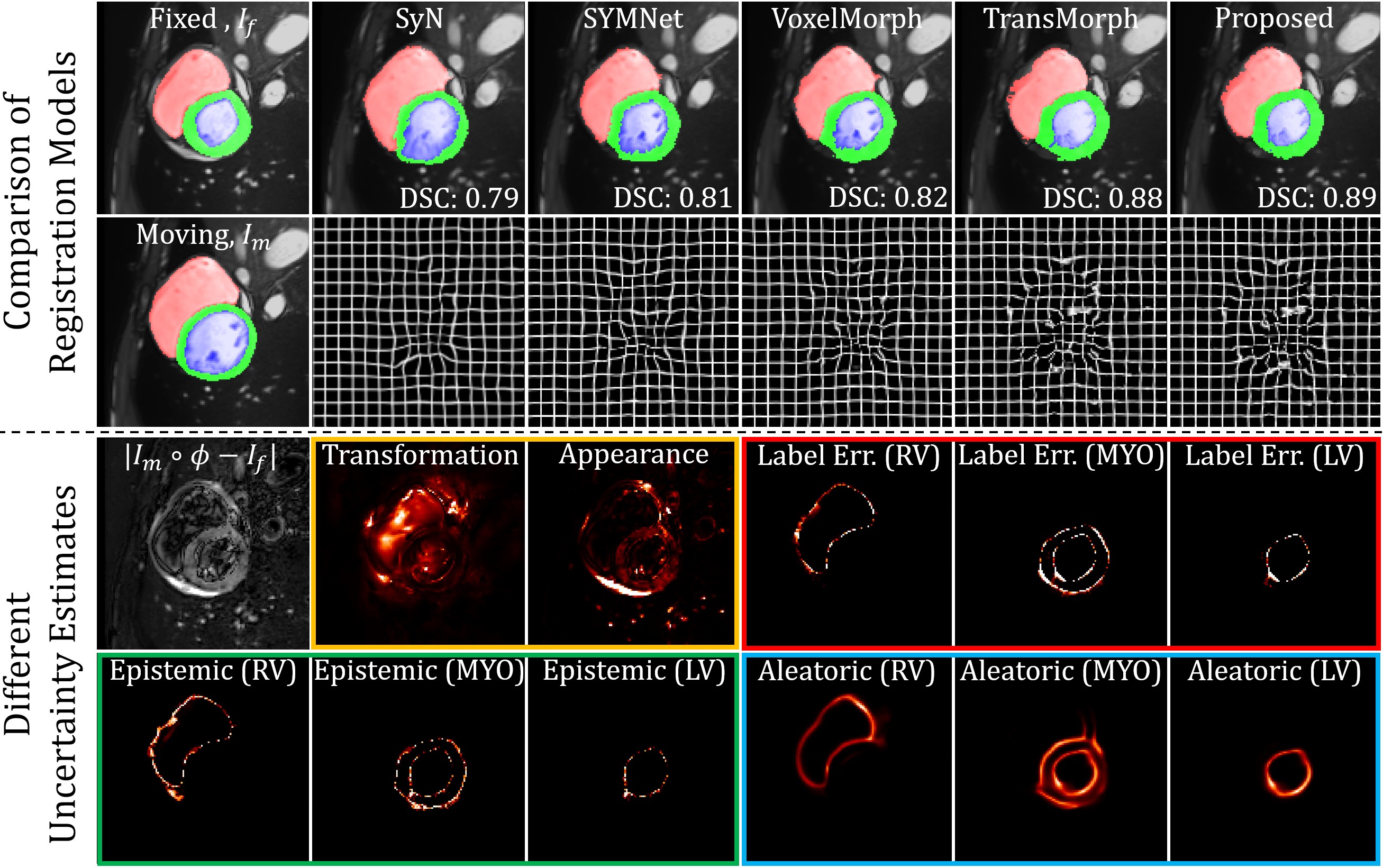}
\caption{Qualitative results of various registration methods, as well as different schemes for registration and segmentation uncertainty quantification using the proposed method. The upper panel shows the qualitative results from different registration methods. The upper panel presents qualitative comparisons across different registration methods. The bottom panel illustrates various uncertainty quantification metrics: the first image depicts the absolute difference in aligned images; yellow highlights registration uncertainties related to transformation and appearance; red indicates label propagation error as squared errors per class; green represents epistemic segmentation uncertainty; and blue delineates aleatoric segmentation uncertainty.} \label{fig:quali_results}
\end{figure*}
\subsection{Epistemic Segmentation Uncertainty}
Let $S_m, S_f: \Omega\to[0,1]^N$ be the $N$-channel anatomical label maps of the moving and fixed image defined over $\Omega$, each channel corresponding to a specific anatomical class. Given the deformation field, $\phi$, the process of warping these 
$N$-channel label maps involves applying \textit{linear interpolation} to each channel. 
Following this interpolation, an argmax operation is performed to obtain the discrete segmentation map. 
The application of linear interpolation on a per-channel basis offers advantages in handling partial volume effects compared to nearest neighbor interpolation.
Here, we define the epistemic segmentation uncertainty as the voxel-wise entropy of the mean of the $T$ propagated labels. 
This method stands apart from the label uncertainty concept introduced in~\cite{luo2019applicability}, which is predicated on the means of occurrence or variance of the $T$ propagated labels. 
Yet, in our context, entropy serves as a more insightful measure for label uncertainty. 
This is because, through linear interpolation, the value in each channel of the interpolated label map at a given voxel effectively acts as a pseudo-probability, indicating the likelihood of that voxel belonging to a certain class. 
This pseudo-probability is maximal (equal to 1) when all surrounding voxels fall within the foreground and decreases when neighboring voxels vary between the foreground and background of the respective class. Consequently, employing entropy in measuring segmentation uncertainty effectively captures the distributional uncertainty across classes.
The mathematical formulation for epistemic segmentation uncertainty is expressed as follows:
\begin{equation}
    U^{Seg.}_{Epi.} = -\sum_{c\in\mathcal{C}}\left(\frac{1}{T}\sum_{i=1}^TS^c_m\circ\phi_i\right)\log\left(\frac{1}{T}\sum_{i=1}^TS^c_m\circ\phi_i\right),
\end{equation}
where $\mathcal{C}\in\{0,1\}$, denotes the background or foreground of each anatomical structure (i.e., channel) as defined in $S_m$. Consequently, $U^{Seg.}_{Epi.}$ comprises $N$ channels, each corresponding to an individual anatomical structure in the label map. Here, $S_{m}^1=S_{m}$, and $S_{m}^0=1-S_{m}$. This approach to estimating epistemic segmentation uncertainty has seen wide application in segmentation DNNs~\cite{eaton2018towards, kendall2017uncertainties, devries2018leveraging, jungo2019assessing}.
However, although we apply the same computational approach used in image segmentation to estimate epistemic segmentation uncertainty for image registration, the resulting uncertainty estimate differs in its underlying principles.
In the context of segmentation DNNs, the epistemic uncertainty captures the variability of a DNN in estimating a pseudo-probability map. Conversely, in registration DNNs, this uncertainty originates from the deformation of a binary label map---in other words, from the way in which the label map is sampled. 
Moreover, to estimate the epistemic segmentation uncertainty for label propagation using image registration requires $S_m$ at test time.
However, this may not always be feasible. In the following subsection, we propose a method to estimate aleatoric segmentation uncertainty. This is done by directly generating it using a compact DNN, alongside the registration DNN.

\subsection{Aleatoric Segmentation Uncertainty}
As depicted in Fig.~\ref{fig:framework}, a compact DNN is conditioned on the appearance difference between the warped and fixed images to estimate the aleatoric segmentation uncertainty associated with the deformation.
This DNN comprises three sequential layers, each consisting of convolution, batch normalization, and Leaky ReLU activation functions, and it concludes with a final convolutional layer.
The inputs to the DNN are the absolute difference between the warped and fixed images (i.e., $\vert I_m\circ\phi - I_f\vert$) and a collection of feature maps from the registration network.
The underlying assumption here is that the error in label propagation is partially reflected in the appearance differences. 
The output is an aleatoric segmentation uncertainty map $( U^{Seg.}_{Ale.} )$, which is of the same size as $S_m$.
Our method, relies on the assumption that the errors between the estimated and the target label map in each channel follow a zero-mean Gaussian distribution. {This assumption is viable due to the use of linear interpolation in warping the labels, which results in the voxel values in each channel of the warped label map being continuous and ranging between [0, 1].} 
Therefore, the aleatoric segmentation uncertainty is essentially the variance of this Gaussian, which we estimate by minimizing a negative log-likelihood~(NLL) loss~\cite{gong2022uncertainty}.
However, it has been demonstrated that the conventional NLL loss could potentially undermine the training of accurate mean predictors~\cite{seitzer2022on}.
To overcome this, we incorporate the $\beta$-NLL~\cite{seitzer2022on}: 
\begin{equation}
\label{Eqn:beta_nll}
\begin{split}
    \mathcal{L}_{\beta-NLL}&=\frac{1}{\Omega}\sum_{p\in\Omega}\floor*{\sigma^{2\beta}(p)}\\
    &\cdot\left(\frac{1}{2}\sigma^{-2}(p)\Vert S_m(p)-S_f(p)\Vert^2+\frac{1}{2}\sigma^2(p)\right),
\end{split}
\end{equation}
where $\floor*{\cdot}$ denotes the stop-gradient operation, and $\beta$ is a hyperparameter that acts to constrain the impact of potentially inaccurate $\sigma^2$ estimates on the loss function. In this study, we set $\beta=1$.
Thus, the aleatoric segmentation uncertainty is represented as: $U^{Seg.}_{Ale.}=\sigma^2$. Notably, estimating $U^{Seg.}_{Ale.}$ necessitates only information from the image domain, making it advantageous when label maps are unavailable during testing.

It is important to underscore that previous research on estimating aleatoric uncertainty for segmentation DNNs~\cite{eaton2018towards, kendall2017uncertainties, devries2018leveraging, jungo2019assessing} generally employs the method introduced by Kendall~\textit{et al.}~\cite{kendall2017uncertainties}, which estimates the aleatoric variance in the logit space. However, this approach is not applicable in the image registration context, as no logits are involved in the label propagation process.
\begin{table*}[t]
\centering
\caption{Quantitative results from the test set. The upper table compares registration performance across various methods, and the lower table compares different approaches to uncertainty quantification by the proposed method.}\label{tab:quant_results}
\fontsize{7.5}{10.5}\selectfont
    \begin{tabular}{ c | c c c c c c}
    % The next line just gives a blank line, to create some space between the caption and table. -AJC
    \multicolumn{7}{c}{}\\[-0.8em]
 \Xhline{1pt}
  \textbf{Method} & \textbf{LV Dice} $\bm{\uparrow}$ & \textbf{RV Dice} $\bm{\uparrow}$ & \textbf{MYO Dice} $\bm{\uparrow}$ & \textbf{Mean Dice} $\bm{\uparrow}$ & \textbf{\%}$\bm{\vert J\vert\leq0\downarrow}$ & \textbf{\%NDV} $\bm{\downarrow}$\\
  \Xhline{1pt}
 \rowcolor{green!20!white}
 Initial & 0.595$\pm$0.162  & 0.608$\pm$0.114 & 0.445$\pm$0.144 & 0.549$\pm$0.112 & - & -\\
 SyN & 0.691$\pm$0.157  & 0.634$\pm$0.134 & 0.687$\pm$0.099 & 0.670$\pm$0.110 & 0.000$\pm$0.001 & 0.000$\pm$0.000\\
 \rowcolor{green!20!white}
 SYMNet & 0.766$\pm$0.111 & 0.797$\pm$0.102& 0.765$\pm$0.060 & 0.776$\pm$0.068 & 1.735$\pm$1.417 &1.627$\pm$1.544\\
 VoxelMorph & 0.836$\pm$0.094 & 0.788$\pm$0.097 & 0.786$\pm$0.058 & 0.803$\pm$0.063 & 0.808$\pm$0.792 & 0.293$\pm$0.328\\
 \rowcolor{green!20!white}
 TransMorph & 0.859$\pm$0.088  & 0.824$\pm$0.093& 0.832$\pm$0.046 & 0.838$\pm$0.057 & 1.216$\pm$0.990 & 0.230$\pm$0.187\\
 \textbf{Proposed} & \textbf{0.861$\pm$0.089} & \textbf{0.824$\pm$0.092} & \textbf{0.834$\pm$0.045} & \textbf{0.839$\pm$0.058} & 1.297$\pm$1.011 & 0.268$\pm$0.206 \\
 \Xhline{1pt}
 & \textbf{LV} $\bm{r~\uparrow}$ & \textbf{RV} $\bm{r~\uparrow}$ & \textbf{MYO} $\bm{r~\uparrow}$ & \textbf{Mean} $\bm{r~\uparrow}$ & &\\
 \Xhline{1pt}
 \rowcolor{green!20!white}
 Transformation & 0.078$\pm$0.044 & 0.107$\pm$0.047 & 0.080$\pm$0.040 & 0.088$\pm$0.030& &\\
 Appearance & 0.053$\pm$0.047 & 0.080$\pm$0.065 & 0.048$\pm$0.029 & 0.060$\pm$0.031& &\\
 \rowcolor{green!20!white}
 Epistemic & 0.531$\pm$0.091 & 0.579$\pm$0.086 & 0.546$\pm$0.046 & 0.552$\pm$0.056& &\\
 Aleatoric & 0.376$\pm$0.084 & 0.371$\pm$0.075 & 0.399$\pm$0.060 & 0.382$\pm$0.059 & &\\
 \rowcolor{green!20!white}
 \textbf{Epi.}+\textbf{Ale.} & \textbf{0.567$\pm$0.073} & \textbf{0.603$\pm$0.077} & \textbf{0.579$\pm$0.035} & \textbf{0.583$\pm$0.045} & &\\
 \Xhline{1pt}
\end{tabular}
\end{table*}
\section{Experiments}
\noindent\textbf{Dataset.} We evaluated the proposed method using two publicly available 3D cardiac MRI datasets: the ACDC challenge~\cite{bernard2018deep} and the M\&Ms challenge~\cite{campello2021multi}. 
These datasets collectively include 470 subjects, with each subject represented by two frames of end-diastolic~(ED) and end-systolic~(ES) stages, along with manually delineated left~(LV) and right ventricle~(RV) blood pools, and the left ventricular myocardium~(MYO).
The subjects were divided into training, validation, and testing sets, consisting of 259, 61, and 150 subjects, respectively. The focus of this study is the registration task between the ED and ES stages.

\noindent\textbf{Evaluation.} We compared our methods with four baselines, one traditional method, SyN~\cite{avants2008symmetric}, and three learning-based methods, VoxelMorph~\cite{balakrishnan2019voxelmorph}, SYMNet~\cite{mok2020fast}, and TransMorph~\cite{chen2022transmorph, chen2022unsupervised}.
To evaluate registration accuracy, we used the Dice coefficient to measure the overlap between the registered anatomical structures. To evaluate the regularity of deformation, we adopted the percentage of all non-positive Jacobian determinants (\%$\vert J\vert\leq0$) and the non-diffeomorphic volume~(\%NDV)~\cite{liu2022finite}. In practice, the estimated segmentation uncertainty should effectively indicate areas where segmentation errors have been made. To evaluate this, we calculated the {Pearson's correlation coefficient~($r$)} between the uncertainty estimate and the label propagation error, which is quantified by the squared error between the propagated and target label maps.

\noindent\textbf{Implementation Details.} Both the proposed method and the baseline methods were trained using the sum of three equally weighted losses: normalized cross-correlation~(NCC), diffusion regularizer~\cite{balakrishnan2019voxelmorph}, and Dice loss.
For the proposed method, we further incorporated Eqn.~\ref{Eqn:beta_nll} as an extra loss function. The models were trained for 500 epochs using the Adam optimizer on an NVIDIA 3090 GPU.

\section{Results}
Figure~\ref{fig:quali_results} shows the qualitative results, with the proposed method demonstrating a better anatomical alignment for the given case.
Regarding uncertainty quantification, it is evident that neither transformation nor appearance registration uncertainty correlates well with the errors in label propagation.
Contrarily, the epistemic segmentation uncertainty demonstrates a robust visual correlation with the error in label propagation, as denoted by ``label err." in Fig.~\ref{fig:quali_results}.
A limitation, however, is the necessity for the label map to be available during test time to estimate such uncertainty. 
Aleatoric segmentation uncertainty, although slightly less correlated, still aligns well with the label propagation error, offering the added benefit of estimation in the absence of label information during testing.
Table~\ref{tab:quant_results} displays the quantitative results, where the proposed method attains the best Dice performance across all anatomical structures.
The results also corroborate that both the epistemic and aleatoric segmentation uncertainty delivered commendable Pearson's $r$ with the label error, while the combination of both types of uncertainties resulted in the highest correlation.

\section{Conclusions}
In this study, we introduced a method to estimate uncertainty in label propagation during image registration.
Rather than altering the registration network, we incorporated an auxiliary compact DNN.
This network is conditioned on the appearance discrepancy between the warped and fixed images, enabling the estimation of both epistemic and aleatoric segmentation uncertainty in addition to the registration uncertainty. 
To our understanding, this represents a pioneering effort to bridge the gap between registration and segmentation uncertainty for image registration.
This fusion provides a comprehensive insight into the registration process in terms of both image warping and label propagation.
We validated the proposed approach using two publicly available datasets, and the results underscore the efficacy of our method.
The proposed uncertainty estimates not only promise to improve atlas-based image segmentation~\cite{aljabar2009multi} by elucidating potential segmentation errors but also hold potential for determining dosimetric uncertainty in cancer therapies associated with the image registration and segmentation processes~\cite{gear2018eanm}.

\section{Acknowledgments}
\label{sec:acknowledgments}
The work was made possible in part by the Johns Hopkins University Discovery Grant~(Co-PI: J.~Chen, Co-PI: A.~Carass).

\bibliographystyle{IEEEbib}
\bibliography{refs}

\end{document}